\documentclass[10pt,english]{article}
\usepackage{times}
\usepackage[T1]{fontenc}
\usepackage[latin1]{inputenc}
\usepackage{array}
\usepackage{amsmath}
\usepackage{amssymb}

\makeatletter

\newcommand{\noun}[1]{\textsc{#1}}
\providecommand{\tabularnewline}{\\}

\usepackage{slashed}

\usepackage{babel}
\makeatother
\begin{document}

\title{\textbf{The Higgs sector of a 3-3-1 model with right-handed neutrinos
to be tested at the LHC}}

\author{\noun{ADRIAN} PALCU}

\date{\emph{Faculty of Exact Sciences - ''Aurel Vlaicu'' University of
Arad, Str. Elena Dr\u{a}goi 2, 310330 - Arad, Romania}}

\maketitle
\begin{abstract}
We explore in this paper certain phenomenological consequences - to
be tested at the LHC - regarding the scalar sector of a $SU(3)_{c}\otimes SU(3)_{L}\otimes U(1)_{X}$
gauge model with right-handed neutrinos. Our analysis is performed
in a particular theoretical approach of treating gauge models with
spontaneous symmetry breaking in which a single free parameter $a$
finally remains to be tuned, once all the Standard Model phenomenology
is recovered. It is also proved that this particular method is flexible
enough as to accommodate the traditional approach in which three VEVs
supply masses for gauge bosons and fermions, while three accompanying
neutral scalars survive the SSB and take part in various interactions.
Two of them exhibit a hierarchy $m(H_{3})\simeq2m(H_{2})$ with masses
below the SM scale $\left\langle \varphi\right\rangle _{SM}=246$
GeV (independently of the parameter $a$) and the third one coming
out very heavy (depending on $a$), at a mass comparable to the overall
breaking scale $\left\langle \varphi\right\rangle $. A plausible
scenario implying $\left\langle \varphi\right\rangle \in1-10$ TeV
is then exploited.


\end{abstract}

\section{Introduction}

The Standard Model (SM) \cite{key-1} - \cite{key-3} - based on the
gauge group $SU(3)_{c}\otimes SU(2)_{L}\otimes U(1)_{Y}$ undergoing
a spontaneous symmetry breakdown (SSB) in its electro-weak sector
- has established itself as a successful theory in explaining the
strong, weak and electromagnetic forces. Nevertheless, some recent
evidences - regarding mainly the neutrino oscillation (see \cite{key-4}
and references therein for an excellent review) - definitely call
for certain extensions of the SM. In order to cover this new and richer
phenomenology, any realistic theoretical model must conceive a consistent
device responsible for generating masses of both fermion and boson
sectors. In the SM this role is accomplished by the so called Higgs
Mechanism \cite{key-5} - \cite{key-9} which - up to date - seems
to be the paradigmatic procedure to give particles their appropriate
masses, while the renormalizability of the model is kept valid. The
Higgs mechanism enforces a suitable SSB up to the electromagnetic
$U(1)_{em}$ group regarded as the residual symmetry of the model.
However, this procedure implies not only a great number of Yukawa
coupling coefficients (undetermined on theoretical ground) in the
fermion sector, but also the existence of a still elusive neutral
scalar particle - namely, the Higgs boson.

Among the possible extensions of the SM, the so called ''3-3-1''
class of models \cite{key-10} - \cite{key-14} emerged two decades
ago and has meanwhile earned a wide reputation through a systematic
and compelling study of its phenomenology. It is based on the $SU(3)_{c}\otimes SU(3)_{L}\otimes U(1)_{X}$gauge
group that undergoes a SSB up to the universal electromagnetic $U(1)_{em}$
symmetry, as in the SM. The discrimination among various models in
this class \cite{key-15} - \cite{key-17} can be done on the particle
content criterion, each model supplying in its own right some new
and distinct phenomenological consequences. We deal here with a particular
model \cite{key-13,key-14} that includes both left-handed and right-handed
neutrinos along with the left-handed charged lepton in triplet representations
of the fermion sector. Besides recovering all the particles coming
from the SM (six quarks and four gauge bosons), it predicts the occurrence
of three new exotic quarks and five new gauge bosons. Apart from other
versions \cite{key-10,key-11} that claim the existence of exotic
electric charges (quarks with $\pm5e/3$, $\pm4e/3$ or bosons with
$\pm2e$), the version under consideration here implies only ordinary
electric charges (even for the exotic particles). 

A few words about the method we have employed to ''solve'' this
class of models. Proposed initially by Cot\emph{\u{a}}escu \cite{key-18},
it essentially consists of a general algebraical procedure in which
electro-weak gauge models with high symmetries ($SU(N)_{L}\otimes U(1)_{X}$)
achieve their SSB in only one step up to $U(1)_{em}$ by means of
a special Higgs mechanism. This supplies a single physical scalar
remaining in the spectrum and the exact expressions for the masses
and neutral currents (charges) of all particles involved in the model.
Here we work out the modified original version and prove that the
procedure can accommodate the traditional approach with three neutral
Higgs scalars surviving the SSB. The proper parametrization of the
scalar sector is paired by an orthogonal restriction among scalar
multiplets that warrants for only three Higgs scalars surviving the
SSB, while all other degrees of freedom (Goldstone bosons) are eaten
by the gauge bosons to become massive. The advantage of this new minimal
Higgs mechanism resides in the fact that a realistic boson mass spectrum
appears to be simply a matter of tuning a single remaining free parameter
$a$. Consequently, the decay widths of these three Higgs scalars
can be expressed in terms of this parameter.

The purpose of this paper is to give an estimate of the properties
of the surviving neutral Higgs bosons from a 3-3-1 model with right-handed
neutrinos (331RHN) based on this particular approach of finally tuning
a single free parameter \cite{key-19,key-20}. We focus especially
on the Higgs bosons couplings such as $HW^{+}W^{-}$, $HZZ$, $HZ^{\prime}Z^{\prime}$,
$HXX^{*}{}$, $HY^{+}Y^{-}$ (where capital letters denote bosons
of the model), in view of obtaining their possible signatures at the
LHC and finally narrowing its mass estimate around the most plausible
values. 

The paper is organized as follows. In Sec.2 we offer a brief review
of the gauge model under consideration here. Possible Higgs boson
decays and other phenomenological consequences are sketched in Sec.3,
while in Sec.4 certain numerical estimates in our scenario are given.
Sec.5 is reserved for sketching our conclusions and suggestions for
experimental search in the Higgs sector at LHC.

\section{Brief review of the model}

The study of the 331RHN models has revealed a rich phenomenology \cite{key-21}
- \cite{key-33} (FCNC processes, $Z\prime$- boson phenomenology,
exotic $T$-quark properties etc.) including some suitable solutions
for the neutrino mass issue \cite{key-34} - \cite{key-40}. With
regard to the scalar sector and Higgs phenomenology a series of papers
\cite{key-41} - \cite{key-44} were published too.

However, we consider it worthwhile presenting the main features of
constructing a 331RHN model. It is based on the gauge group $SU(3)_{c}\otimes SU(3)_{L}\otimes U(1)_{X}$
and the main pieces are the irreducible representations which correspond
to fermion left-handed multiplets. The fermion content is the following:

\subparagraph*{Lepton families}

\begin{equation}
\begin{array}{ccccc}
f_{\alpha L}=\left(\begin{array}{c}
\nu_{\alpha}^{c}\\
\nu_{\alpha}\\
e_{\alpha}\end{array}\right)_{L}\sim(\mathbf{1,3},-1/3) &  &  &  & e_{\alpha R}\sim(\mathbf{1},\mathbf{1},-1)\end{array}\label{Eq.1}\end{equation}

\subparagraph*{Quark families}

\begin{equation}
\begin{array}{ccc}
Q_{iL}=\left(\begin{array}{c}
D_{i}\\
-d_{i}\\
u_{i}\end{array}\right)_{L}\sim(\mathbf{3,3^{*}},0) &  & Q_{3L}=\left(\begin{array}{c}
U_{3}\\
u_{3}\\
d_{3}\end{array}\right)_{L}\sim(\mathbf{3},\mathbf{3},+1/3)\end{array}\label{Eq.2}\end{equation}

\begin{equation}
\begin{array}{ccc}
d_{iR},d_{3R}\sim(\mathbf{3},\mathbf{1},-1/3) &  & u_{iR},u_{3R}\sim(\mathbf{3},\mathbf{1},+2/3)\end{array}\label{Eq.3}\end{equation}

\begin{equation}
\begin{array}{ccccccccc}
U_{3R}\sim(\mathbf{3,1},+2/3) &  &  &  &  &  &  &  & D_{iR}\sim(\mathbf{3,1},-1/3)\end{array}\label{Eq.4}\end{equation}
with $i=1,2$. 

In the representations displayed above one has to assume that two
generations of quarks transform differently from the third one in
order to cancel all the axial anomalies (by an interplay between families,
although each one remains anomalous by itself). In this way one prevents
the model from compromising its renormalizability by triangle diagrams.
The capital letters denote the exotic quarks included in each family.
Many authors consider that $U_{3R}=T$ and $D_{iR}=D,S$ as a possible
explanation of the unusual heavy masses of the third generation of
quarks, but we restrict ourselves here to make no particular choice.

\subparagraph*{Gauge bosons}

The gauge bosons of the model are connected to the generators of the
$su(3)$ Lie algebra, expressed by the usual Gell-Mann matrices $T_{a}=\lambda_{a}/2$
. So, the Hermitian diagonal generators of the Cartan sub-algebra
are \begin{equation}
D_{1}=T_{3}=\frac{1}{2}{\textrm{Diag}}(1,-1,0)\,,\quad D_{2}=T_{8}=\frac{1}{2\sqrt{3}}\,{\textrm{Diag}}(1,1,-2)\,.\label{Eq,5}\end{equation}
 In this basis the gauge fields are $A_{\mu}^{0}$ (corresponding
to the Lie algebra of the group $U(1)_{X}$) and $A_{\mu}\in su(3)$,
that can be put as \begin{equation}
A_{\mu}=\frac{1}{2}\left(\begin{array}{ccc}
A_{\mu}^{3}+A_{\mu}^{8}/\sqrt{3} & \sqrt{2}X_{\mu} & \sqrt{2}Y_{\mu}\\
\\\sqrt{2}X_{\mu}^{*} & -A_{\mu}^{3}+A_{\mu}^{8}/\sqrt{3} & \sqrt{2}W_{\mu}\\
\\\sqrt{2}Y_{\mu}^{*} & \sqrt{2}W_{\mu}^{*} & -2A_{\mu}^{8}/\sqrt{3}\end{array}\right),\label{Eq.6}\end{equation}
 where $\sqrt{2}W_{\mu}^{\pm}=A_{\mu}^{6}\mp iA_{\mu}^{7}$, $\sqrt{2}Y_{\mu}^{\pm}=A_{\mu}^{4}\pm iA_{\mu}^{5}$,
and $\sqrt{2}X_{\mu}=A_{\mu}^{1}-iA_{\mu}^{2}$, respectively. One
notes that apart from the charged Weinberg bosons ($W^{\pm}$) from
SM, there are two new complex boson fields, $X$ (neutral) and $Y$
(charged). 

The diagonal Hermitian generators are associated to the neutral gauge
bosons $A_{\mu}^{em}$, $Z_{\mu}$and $Z_{\mu}^{\prime}$. On the
diagonal terms in Eq.(\ref{Eq.6}) a generalized Weinberg transformation
(gWt) must be performed in order to consequently separate the massless
electromagnetic field from the other two neutral massive fields. The
details of this procedure can be found in Ref. \cite{key-18} and
its concrete realization in the model of interest here in Refs. \cite{key-19,key-20}.

\section{Scalar sector}

In the general method \cite{key-18}, the scalar sector of any gauge
model must consist of $n$ Higgs multiplets $\phi^{(1)}$, $\phi^{(2)}$,
... , $\phi^{(n)}$ satisfying the orthogonal condition $\phi^{(i)+}\phi^{(j)}=\varphi^{2}\delta_{ij}$
in order to eliminate unwanted Goldstone bosons that could survive
the SSB. Here, $\varphi$ is a gauge-invariant real field variable
acting as a norm in the scalar space and $n$ is the dimension of
the fundamental irreducible representation of the gauge group. The
parameter matrix $\eta=\left(\eta_{0},\eta{}_{1},\eta{}_{2}..,\eta{}_{n}\right)$
with the property $Tr\eta^{2}=1-\eta_{0}^{2}$ is a key ingredient
of the method: it is introduced in order to obtain a non-degenerate
boson mass spectrum. Obviously, $\eta_{0},\eta{}_{i}\in[0,1)$. Then,
the Higgs Lagrangian density (Ld) reads:

\begin{equation}
\mathcal{L}_{H}=\frac{1}{2}\eta_{0}^{2}\partial_{\mu}\phi\partial^{\mu}\phi+\frac{1}{2}\sum_{i=1}^{n}\eta_{i}^{2}\left(D_{\mu}\phi^{(i)}\right)^{+}\left(D^{\mu}\phi^{(i)}\right)-V(\phi^{(i)})\label{Eq. 7}\end{equation}
where $D_{\mu}\phi^{(i)}=\partial_{\mu}\phi^{(i)}-i(gA_{\mu}+g^{\prime}y^{(i)}A_{\mu}^{0})\phi^{(i)}$
act as covariant derivatives of the model, and $g$ and $g^{\prime}$
the coupling constants of the groups $SU(N)_{L}$ and $U(1)_{X}$
respectively. Real characters $y^{(i)}$ stand as a kind of hyper-charge
of the new theory. 

For the particular 331RHN model under consideration here the most
general choice of parameters is given by the matrix $\eta^{2}=\left(1-\eta_{0}^{2}\right)Diag\left[1-a,\frac{1}{2}\left(a-b\right),\frac{1}{2}\left(a+b\right)\right]$.
It obviously meets the trace condition required by the general method
for any $a,b\in[0,1)$. After imposing the phenomenological condition
$M_{Z}^{2}=M_{W}^{2}/\cos^{2}\theta_{W}$ (confirmed at the SM level)
the procedure of diagonalizing the neutral boson mass matrix \cite{key-19,key-20}
eliminates a parameter and thus the parameter matrix becomes $\eta^{2}=\left(1-\eta_{0}^{2}\right)Diag\left[1-a,a\frac{\left(1-\tan^{2}\theta_{W}\right)}{2},a\frac{1}{2\cos^{2}\theta_{W}}\right]$.

\subsection{Scalar fields redefinition }

In the following we accommodate our method with the traditional approach
in which there are 3 distinct VEVs resulting from the potential minimum
condition. For this purpose we redefine the scalar triplets as following

\begin{equation}
\phi^{(1)}\rightarrow\eta{}_{1}\phi^{(1)}\equiv\rho\,,\quad\phi^{(2)}\rightarrow\eta{}_{2}\phi^{(2)}\equiv\chi\,,\quad\phi^{(3)}\rightarrow\eta{}_{3}\phi^{(3)}\equiv\phi\,.\label{Eq, 8}\end{equation}
or in an equivalent notation (with the upper index showing the electric
charge of the filed it labels):

\begin{equation}
\rho=\left(\begin{array}{c}
\rho^{0}\\
\\\rho^{0}\\
\\\rho^{-}\end{array}\right)\,,\quad\chi=\left(\begin{array}{c}
\chi^{0}\\
\\\chi^{0}\\
\\\chi^{-}\end{array}\right)\,,\quad\phi=\left(\begin{array}{c}
\phi^{+}\\
\\\phi^{+}\\
\\\phi^{0}\end{array}\right)\,.\label{Eq. 9}\end{equation}

Obviously, these new fields obey orthogonal relations in a new form,
namely:

\begin{equation}
\rho^{+}\rho=\eta_{1}^{2}\varphi^{2}\,,\quad\chi^{+}\chi=\eta_{2}^{2}\varphi^{2}\,,\quad\phi^{+}\phi=\eta_{3}^{2}\varphi^{2}\,.\label{Eq, 10}\end{equation}

The simplest potential that preserves renormalizability can be put
now in the following form:

\begin{equation}
\begin{array}{ccl}
V & = & -\mu_{1}^{2}\rho^{+}\rho-\mu_{2}^{2}\chi^{+}\chi-\mu_{3}^{2}\phi^{+}\phi+\lambda_{1}\left(\rho^{+}\rho\right)^{2}+\lambda_{2}\left(\chi^{+}\chi\right)^{2}+\lambda_{3}\left(\phi^{+}\phi\right)^{2}\\
\\ &  & +\lambda_{4}\left(\rho^{+}\rho\right)\left(\chi^{+}\chi\right)+\lambda_{5}\left(\rho^{+}\rho\right)\left(\phi^{+}\phi\right)+\lambda_{6}\left(\phi^{+}\phi\right)\left(\chi^{+}\chi\right).\end{array}\label{Eq. 11}\end{equation}

One can easily observe that the SSB is accomplished in the unitary
gauge by three VEVs, as follows:

\begin{equation}
\left(\begin{array}{c}
\eta_{1}\left\langle \varphi\right\rangle +H_{\rho}\\
\\0\\
\\0\end{array}\right)\,,\quad\left(\begin{array}{c}
0\\
\\\eta_{2}\left\langle \varphi\right\rangle +H_{\chi}\\
\\0\end{array}\right)\,,\quad\left(\begin{array}{c}
0\\
\\0\\
\\\eta_{3}\left\langle \varphi\right\rangle +H_{\phi}\end{array}\right)\,.\label{Eq. 12}\end{equation}
with 

\begin{equation}
\left\langle \varphi\right\rangle =\frac{\sqrt{\mu_{1}^{2}\eta_{1}^{2}+\mu_{2}^{2}\eta_{2}^{2}+\mu_{3}^{2}\eta_{3}^{2}}}{\sqrt{2\left(\lambda_{1}\eta_{1}^{4}+\lambda_{2}\eta_{2}^{4}+\lambda_{3}\eta_{3}^{4}\right)+\lambda_{4}\eta_{1}^{2}\eta_{2}^{2}+\lambda_{5}\eta_{1}^{2}\eta_{3}^{2}+\lambda_{6}\eta_{2}^{2}\eta_{3}^{2}}}\label{Eq. 13}\end{equation}
resulting from the minimum condition applied to the above potential
(\ref{Eq. 11}). 

$H_{1}$, $H_{2}$, $H_{3}$ are the physical Higgs fields surviving
the SSB. Let's look for their couplings. To this end one can write
explicitly the terms in the potential $V$ after SSB took place:

\begin{equation}
\begin{array}{cll}
V & = & -\left[\mu_{1}^{2}\left(\eta_{1}\left\langle \varphi\right\rangle +H_{\rho}\right)^{2}+\mu_{2}^{2}\left(\eta_{2}\left\langle \varphi\right\rangle +H_{\chi}\right)^{2}+\mu_{3}^{2}\left(\eta_{3}\left\langle \varphi\right\rangle +H_{\phi}\right)^{2}\right]\\
\\ &  & +\left[\lambda_{1}\left(\eta_{1}\left\langle \varphi\right\rangle +H_{\rho}\right)^{4}+\lambda_{2}\left(\eta_{2}\left\langle \varphi\right\rangle +H_{\chi}\right)^{4}+\lambda_{3}\left(\eta_{3}\left\langle \varphi\right\rangle +H_{\phi}\right)^{4}\right]\\
\\ &  & +\left[\lambda_{4}\left(\eta_{1}\left\langle \varphi\right\rangle +H_{\rho}\right)^{2}\left(\eta_{2}\left\langle \varphi\right\rangle +H_{\chi}\right)^{2}+\lambda_{5}\left(\eta_{1}\left\langle \varphi\right\rangle +H_{\rho}\right)^{2}\left(\eta_{3}\left\langle \varphi\right\rangle +H_{\phi}\right)^{2}\right]\\
\\ &  & +\lambda_{6}\left(\eta_{2}\left\langle \varphi\right\rangle +H_{\chi}\right)^{2}\left(\eta_{3}\left\langle \varphi\right\rangle +H_{\phi}\right)^{2}.\end{array}\label{Eq. 14}\end{equation}

\subsection{Scalar fields couplings}

The next step is to identify for each Higgs its own coupling terms.
These are in order.

\paragraph*{(i) linear terms \textmd{(must be absent - as in the SM - so one
gets three constraints on the parameters):}}

\[
\begin{array}{cc}
H_{\rho}: & -\mu_{1}^{2}+\left(2\lambda_{1}\eta_{1}^{2}+\lambda_{4}\eta_{2}^{2}+\lambda_{5}\eta_{3}^{2}\right)\left\langle \varphi\right\rangle ^{2}\end{array}=0\]

\begin{equation}
\begin{array}{cc}
H_{\chi}: & -\mu_{2}^{2}+\left(2\lambda_{2}\eta_{2}^{2}+\lambda_{4}\eta_{1}^{2}+\lambda_{6}\eta_{3}^{2}\right)\left\langle \varphi\right\rangle ^{2}\end{array}=0\label{Eq. 15}\end{equation}

\[
\begin{array}{cc}
H_{\phi}: & -\mu_{3}^{2}+\left(2\lambda_{3}\eta_{3}^{2}+\lambda_{5}\eta_{1}^{2}+\lambda_{6}\eta_{2}^{2}\right)\left\langle \varphi\right\rangle ^{2}\end{array}=0\]

\paragraph*{(ii) mass terms: }

\[
\begin{array}{cc}
H_{\rho}H_{\rho}: & -\mu_{1}^{2}\end{array}+\left(6\lambda_{1}\eta_{1}^{2}+\lambda_{4}\eta_{2}^{2}+\lambda_{5}\eta_{3}^{2}\right)\left\langle \varphi\right\rangle ^{2}=4\lambda_{1}\eta_{1}^{2}\left\langle \varphi\right\rangle ^{2}\]

\begin{equation}
\begin{array}{cc}
H_{\chi}H_{\chi}: & -\mu_{2}^{2}\end{array}+\left(6\lambda_{2}\eta_{2}^{2}+\lambda_{4}\eta_{1}^{2}+\lambda_{6}\eta_{3}^{2}\right)\left\langle \varphi\right\rangle ^{2}=4\lambda_{2}\eta_{2}^{2}\left\langle \varphi\right\rangle ^{2}\label{Eq. 16}\end{equation}

\[
\begin{array}{cc}
H_{\phi}H_{\phi}: & -\mu_{3}^{2}\end{array}+\left(6\lambda_{3}\eta_{3}^{2}+\lambda_{5}\eta_{1}^{2}+\lambda_{6}\eta_{2}^{2}\right)\left\langle \varphi\right\rangle ^{2}=4\lambda_{3}\eta_{3}^{2}\left\langle \varphi\right\rangle ^{2}\]

\begin{equation}
\begin{array}{c}
H_{\rho}H_{\chi}:4\lambda_{4}\eta_{1}\eta_{2}\left\langle \varphi\right\rangle ^{2}\,,\,\end{array}H_{\rho}H_{\phi}:4\lambda_{5}\eta_{1}\eta_{3}\left\langle \varphi\right\rangle ^{2}\,,\, H_{\phi}H_{\chi}:4\lambda_{6}\eta_{2}\eta_{3}\left\langle \varphi\right\rangle ^{2}\,.\label{Eq. 17}\end{equation}

\paragraph*{(iii) $HHH$ trilinear terms:}

\begin{equation}
\begin{array}{ccccc}
H_{\rho}H_{\rho}H_{\rho}:\lambda_{1}\eta_{1}\left\langle \varphi\right\rangle  & , & H_{\rho}H_{\chi}H_{\chi}:2\lambda_{4}\eta_{1}\left\langle \varphi\right\rangle  & , & H_{\rho}H_{\phi}H_{\phi}:2\lambda_{5}\eta_{1}\left\langle \varphi\right\rangle \,,\\
\\H_{\chi}H_{\rho}H_{\rho}:2\lambda_{4}\eta_{2}\left\langle \varphi\right\rangle  & , & H_{\chi}H_{\chi}H_{\chi}:4\lambda_{2}\eta_{2}\left\langle \varphi\right\rangle  & , & H_{\chi}H_{\phi}H_{\phi}:2\lambda_{6}\eta_{2}\left\langle \varphi\right\rangle \,,\\
\\H_{\phi}H_{\rho}H_{\rho}:2\lambda_{5}\eta_{3}\left\langle \varphi\right\rangle  & , & H_{\phi}H_{\chi}H_{\chi}:2\lambda_{6}\eta_{3}\left\langle \varphi\right\rangle  & , & H_{\phi}H_{\phi}H_{\phi}:4\lambda_{3}\eta_{3}\left\langle \varphi\right\rangle \,.\end{array}\label{Eq. 18}\end{equation}

\paragraph*{(iv) $HHHH$ quartic terms:}

\begin{equation}
\begin{array}{ccccc}
H_{\rho}H_{\rho}H_{\rho}H_{\rho}:\lambda_{1} & , & H_{\chi}H_{\chi}H_{\chi}H_{\chi}:\lambda_{2} & , & H_{\phi}H_{\phi}H_{\phi}H_{\phi}:\lambda_{3}\,.\end{array}\label{Eq. 19}\end{equation}

\subsection{Higgs masses}

From the above expressions one can identify the Higgs mass matrix
as:

\begin{equation}
M_{H}^{2}=4\left(\begin{array}{ccccc}
\lambda_{1}\eta_{1}^{2} &  & \lambda_{4}\eta_{1}\eta_{2} &  & \lambda_{5}\eta_{1}\eta_{3}\\
\\\lambda_{4}\eta_{1}\eta_{2} &  & \lambda_{2}\eta_{2}^{2} &  & \lambda_{6}\eta_{2}\eta_{3}\\
\\\lambda_{5}\eta_{1}\eta_{3} &  & \lambda_{6}\eta_{2}\eta_{3} &  & \lambda_{3}\eta_{3}^{2}\end{array}\right)\left\langle \varphi\right\rangle ^{2}\label{Eq. 20}\end{equation}

In the phenomenological case of interest here, as we will see in Sec.4,
$\left\langle \rho\right\rangle \gg\left\langle \chi\right\rangle ,\left\langle \phi\right\rangle $
that is $\eta_{1}\rightarrow1$ and $\eta_{2},\eta_{3}\rightarrow0$
in our parametrization, in order to ensure a correct boson mass spectrum
\cite{key-19,key-20}. Consequently, the Higgs mass matrix can be
computed by eluding the very small entries in its texture and considering
the mass of the first Higgs boson - $H_{1}\cong H_{\rho}$ - as:

\begin{equation}
m_{1}^{2}\cong4\lambda_{1}\eta_{1}^{2}\left\langle \varphi\right\rangle ^{2}\label{Eq. 21}\end{equation}

Assuming this Higgs ($H_{1}$) does not mix with the two remaining
ones, their physical; basis can be reached by a simple $2\times2$
rotation: 

\begin{equation}
H_{2}\cong\frac{\lambda_{5}\eta_{2}H_{\chi}-\lambda_{4}\eta_{3}H_{\phi}}{\sqrt{\lambda_{4}^{2}\eta_{3}^{2}+\lambda_{5}^{2}\eta_{2}^{2}}}\label{Eq. 22}\end{equation}

\begin{equation}
H_{3}\cong\frac{\lambda_{4}\eta_{3}H_{\chi}+\lambda_{5}\eta_{2}H_{\phi}}{\sqrt{\lambda_{4}^{2}\eta_{3}^{2}+\lambda_{5}^{2}\eta_{2}^{2}}}\label{Eq. 23}\end{equation}
Hence, their corresponding masses are:

\begin{equation}
m_{2}^{2}\cong2\eta_{2}^{2}\left(\frac{\lambda_{3}\lambda_{4}-\lambda_{5}\lambda_{6}}{\lambda_{4}}\right)\left\langle \varphi\right\rangle ^{2}\label{Eq. 24}\end{equation}

\begin{equation}
m_{2}^{2}\cong2\left(\lambda_{3}\eta_{2}^{2}+\frac{\lambda_{4}\lambda_{6}}{\lambda_{5}}\eta_{3}^{2}\right)\left\langle \varphi\right\rangle ^{2}\label{Eq. 25}\end{equation}

For the sake of simplicity here is the point where one can make certain
assumptions, namely considering, $\lambda_{1}\simeq\lambda_{2}\simeq\lambda_{3}\equiv\lambda$
and $\lambda_{4}\simeq\lambda_{5}\simeq\lambda_{6}\equiv\lambda^{\prime}.$
By inserting these notations into Eqs. (\ref{Eq. 20}), (\ref{Eq. 25})
and (\ref{Eq. 26}) one can get the following expressions: 

\begin{equation}
m_{1}^{2}\cong4\lambda\eta_{1}^{2}\left\langle \varphi\right\rangle ^{2}\label{Eq. 26}\end{equation}

\begin{equation}
m_{2}^{2}\cong2\eta_{2}^{2}\left(\lambda-\lambda^{\prime}\right)\left\langle \varphi\right\rangle ^{2}\label{Eq. 27}\end{equation}

\begin{equation}
m_{3}^{2}\cong2\left(\lambda\eta_{2}^{2}+\lambda^{\prime}\eta_{3}^{2}\right)\left\langle \varphi\right\rangle ^{2}\label{Eq. 28}\end{equation}

Obviously $\lambda^{\prime}$ has to range in $[0,\lambda)$ in order
to keep meaningful the whole procedure of identifying Higgs masses.
We roughly inspect three cases, accounting certain particular values
of the ratio $\lambda^{\prime}/\lambda\,:\,.0.,\,\frac{1}{2},\,.1.$

The heaviest Higgs gets in all three cases its mass: as $m_{1}\cong2\sqrt{\lambda\left(1-a\right)}\left\langle \varphi\right\rangle $.

\paragraph*{Case 1: \textmd{If $\lambda^{\prime}=0$ one gets $m_{2}=m_{3}\cong\sqrt{\lambda a\left(1-\tan^{2}\theta_{W}\right)}\left\langle \varphi\right\rangle $,
two small but degenerate masses for the lighter Higgs bosons. This
setting is less probable since it means that there are suppressed
quartic terms like $HHHH$. }}

\paragraph*{Case 2: \textmd{If $\lambda^{\prime}=\lambda$ one gets $m_{2}=0$
and $m_{3}\cong\sqrt{2\lambda a}\left\langle \varphi\right\rangle $.
This setting also has to be ruled out, since a massless Higgs which
couples to SM bosons causes logarithmically divergent contributions
in 1-loop corrections to $\rho$ parameter and W boson mass, spoiling
thus the renormalizability of the model. }}

\paragraph*{Case 3: \textmd{If $\lambda^{\prime}=\lambda/2$ some plausible numerical
estimates can be performed. First of all, $H_{3}$ can be seen as
the SM-like Higgs boson. Since the custodial symmetry of the SM is
no more valid here, the second SM-like Higgs doublet is missing, so
that the new $H_{2}$ takes the role of giving quarks their masses.
$H_{1}$and $H_{2}$ are the new Higgs bosons specific to this 331RHN
model. The three masses are:} }

\begin{equation}
m_{1}\cong2\sqrt{\lambda\left(1-a\right)}\left\langle \varphi\right\rangle \label{Eq. 29}\end{equation}

\begin{equation}
m_{2}\cong\sqrt{\frac{1}{2}\lambda a\left(1-\tan^{2}\theta_{W}\right)}\left\langle \varphi\right\rangle \label{Eq. 30}\end{equation}

\begin{equation}
m_{3}\cong\sqrt{\frac{1}{2}\lambda a\left(\frac{4\cos^{2}\theta_{W}-1}{\cos^{2}\theta_{W}}\right)}\left\langle \varphi\right\rangle \label{Eq. 31}\end{equation}

The resulting expressions for Higgs masses in Case 3 suggest that
$m_{2}\simeq2m_{3}$ (both are in quite the same range - the SM scale
- since $\sqrt{\left(1-\tan^{2}\theta_{W}\right)}\cong0.845$and $\sqrt{\left(\frac{4\cos^{2}\theta_{W}-1}{\cos^{2}\theta_{W}}\right)}\cong1.65$
for $\sin^{2}\theta_{W}\cong0.223$ \cite{key-45}), and $m_{1}$lies
in TeV domain, as it will be seen more clearly in the next section,
when the parameters will be properly tuned.

\subsection{Higgs interactions}

In order to analyze the possible phenomenological consequences regarding
the Higgs sector and its likely processes (decays, pair production
etc) one has to observe the terms that provide us with the couplings
of the physical Higgs bosons to the gauge bosons of the model (HBB).
They can be read from the resulting Ld in unitary gauge after SSB,
namely:

\begin{equation}
\begin{array}{ccl}
\mathcal{L} & = & \frac{g^{2}}{4}\left[\left(\eta_{1}\left\langle \varphi\right\rangle +H_{\rho}\right)^{2}+\left(\eta_{2}\left\langle \varphi\right\rangle +H_{\chi}\right)^{2}\right]X_{\mu}^{+}X^{\mu}\\
\\ & + & \frac{g^{2}}{4}\left[\left(\eta_{1}\left\langle \varphi\right\rangle +H_{\rho}\right)^{2}+\left(\eta_{3}\left\langle \varphi\right\rangle +H_{\phi}\right)^{2}\right]Y_{\mu}^{+}Y^{\mu}\\
\\ & + & \frac{g^{2}}{4}\left[\left(\eta_{2}\left\langle \varphi\right\rangle +H_{\chi}\right)^{2}+\left(\eta_{3}\left\langle \varphi\right\rangle +H_{\phi}\right)^{2}\right]W_{\mu}^{+}W^{\mu}\\
\\ & + & \frac{g^{2}}{8\cos^{2}\theta_{W}}\left[\left(\eta_{2}\left\langle \varphi\right\rangle +H_{\chi}\right)^{2}+\left(\eta_{3}\left\langle \varphi\right\rangle +H_{\phi}\right)^{2}\right]Z_{\mu}Z^{\mu}\\
\\ & + & \frac{g^{2}}{8}\left(\frac{4\cos^{2}\theta_{W}}{3-4\sin^{2}\theta_{W}}\right)\left(\eta_{1}\left\langle \varphi\right\rangle +H_{\rho}\right)^{2}Z_{\mu}^{\prime}Z^{\prime\mu}\\
\\ & + & \frac{g^{2}}{8}\frac{\left(1-2\sin^{2}\theta_{W}\right)^{2}}{\left(3-4\sin^{2}\theta_{W}\right)\cos^{2}\theta_{W}}\left(\eta_{2}\left\langle \varphi\right\rangle +H_{\chi}\right)Z_{\mu}^{\prime}Z^{\prime\mu}\\
\\ & + & \frac{g^{2}}{8}\frac{1}{\left(3-4\sin^{2}\theta_{W}\right)\cos^{2}\theta_{W}}\left(\eta_{3}\left\langle \varphi\right\rangle +H_{\phi}\right)Z_{\mu}^{\prime}Z^{\prime\mu}.\end{array}\label{Eq. 32}\end{equation}

\subsubsection{Boson mass spectrum}

From the above expression the boson mass spectrum can be inferred,
by simply identifying the proper terms as the mass Ld:

\begin{equation}
\begin{array}{ccl}
\mathcal{L}_{mass} & = & (2M_{W}^{2}W_{\mu}^{+}W^{\mu}+M_{Z}^{2}Z_{\mu}Z^{\mu}\\
\\ & + & 2M_{X}^{2}X_{\mu}^{+}X^{\mu}+2M_{Y}^{2}Y_{\mu}^{+}Y^{\mu}+M_{Z^{\prime}}^{2}Z_{\mu}^{\prime}Z^{\prime\mu}).\end{array}\label{Eq. 33}\end{equation}

A rapid calculus drives straightforwardly from Ld \ref{Eq. 32} to
the boson mass spectrum previously obtained with our method in Refs.
\cite{key-19,key-20}, namely:

\begin{itemize}
\item $M_{W}^{2}=m^{2}a$ 
\item $M_{Y}^{2}=m^{2}\left(1-a/2\cos^{2}\theta_{W}\right)$ 
\item $M_{X}^{2}=m^{2}\left[1-a(1-\tan^{2}\theta_{W})/2\right]$ 
\item $M_{Z}^{2}=m^{2}a/\cos^{2}\theta_{W}$ 
\item $M_{Z^{\prime}}^{2}=m^{2}\left[4\cos^{2}\theta_{W}-a\left(3-4\sin^{2}\theta_{W}+\tan^{2}\theta_{W}\right)\right]/\left(3-4\sin^{2}\theta_{W}\right)$
\end{itemize}
We have made the notation: $m^{2}=g^{2}\left\langle \varphi\right\rangle ^{2}(1-\eta_{0}^{2})/4$.
The mass scale is now just a matter of tuning the parameter $a$ in
accordance with the possible values for $\left\langle \varphi\right\rangle $.
One can set parameter $\eta_{0}^{2}$ (of the original method) very
small so that, for our purpose here, $m^{2}\simeq g^{2}\left\langle \varphi\right\rangle ^{2}/4$. 

One can note for the neutral bosons sector that the diagonalization
of the resulting mass matrix \cite{key-19} has been performed by
imposing the specific relation between $M_{W}$ and $m_{Z}$, namely
$M_{Z}^{2}=M_{W}^{2}/\cos^{2}\theta_{W}$. That is why one finally
remains with a single free parameter to be tuned $a$. Moreover, the
rotation matrix doing the diagonalization job has established the
mixing angle $\sin\phi=1/2\sqrt{1-\sin^{2}\theta_{W}}$. The traditional
approach in the literature assumes $\phi$ as a free parameter restricted
on experimental ground. Here it is fixed, the role of ensuring the
experimentally observed gap between $m(Z^{\prime})$ and $m(Z)$ being
realized exclusively by the free parameter $a$. In addition, we mention
that the correct coupling match is recovered through our method, namely
$g^{\prime}=g\sqrt{3}\sin\theta_{W}/\sqrt{3-4\sin^{2}\theta_{W}}$.
All the couplings in the neutral currents of the model (or, in other
words, the neutral charges of the fermions) are exactly obtained and
need no approximation. They also reproduce for the SM fermions their
established values (for the detailed list, the reader is referred
to the Table in Ref. \cite{key-20}).

\subsubsection{Higgs fields couplings}

From (\ref{Eq. 32}) combined with Eqs. (\ref{Eq. 22}) - (\ref{Eq. 23})
one can get the $HBB$ couplings for the real Higgs fields. Their
general expressions are put in the first two columns of the Table
1, while their numerical values in the scenario considered in Sec.4
are displayed in the last column of the same Table1. 

\begin{equation}
g\left(H_{1}BB\right)\simeq g\left(H_{\rho}BB\right)\label{Eq. 34}\end{equation}

\begin{equation}
g\left(H_{2}BB\right)\simeq\left[g\left(H_{\chi}BB\right)\sqrt{\frac{1-\tan^{2}\theta_{W}}{2}}-g\left(H_{\phi}BB\right)\sqrt{\frac{1}{2\cos^{2}\theta_{W}}}\right]\label{Eq. 35}\end{equation}

\begin{equation}
g\left(H_{3}BB\right)\simeq\left[g\left(H_{\chi}BB\right)\sqrt{\frac{1}{2\cos^{2}\theta_{W}}}+g\left(H_{\phi}BB\right)\sqrt{\frac{1-\tan^{2}\theta_{W}}{2}}\right]\label{Eq. 36}\end{equation}

The couplings of the form $HHBB$ can be obtained from the ones in
Eqs. (\ref{Eq. 35}) - (\ref{Eq. 36}) by simply dividing by $2\left\langle \varphi\right\rangle $. 

\begin{table}

\caption{HBB couplings}

\begin{tabular}{l|cl|clc}
\hline 
Couplings HBB&
&
 $\times$($m^{2}/\left\langle \varphi\right\rangle $)&
&
$\times(2M_{W}^{2}/\left\langle \varphi\right\rangle _{SM})$&
\tabularnewline
\hline
\hline 
&
&
&
&
&
\tabularnewline
$H_{1}X_{\mu}^{+}X^{\mu}$&
&
$2\eta_{1}$&
&
$\sqrt{\frac{1-a}{a}}$&
\tabularnewline
&
&
&
&
&
\tabularnewline
$H_{1}Y_{\mu}^{+}Y^{\mu}$&
&
$2\eta_{1}$&
&
$\sqrt{\frac{1-a}{a}}$&
\tabularnewline
&
&
&
&
&
\tabularnewline
$H_{1}Z_{\mu}^{\prime}Z^{\prime\mu}$&
&
$\frac{4\cos^{2}\theta_{W}}{3-4\sin^{2}\theta_{W}}\eta_{1}$&
&
$\left(\frac{1}{2}\right)$$\frac{4\cos^{2}\theta_{W}}{3-4\sin^{2}\theta_{W}}\sqrt{\frac{1-a}{a}}=0.73\sqrt{\frac{1-a}{a}}$&
\tabularnewline
&
&
&
&
&
\tabularnewline
$H_{2}X_{\mu}^{+}X^{\mu}$&
&
$2\eta_{2}^{2}\frac{1}{\sqrt{a}}$&
&
$\left(\frac{1}{2}\right)\left(1-\tan^{2}\theta_{W}\right)=0.36$&
\tabularnewline
&
&
&
&
&
\tabularnewline
$H_{2}Y_{\mu}^{+}Y^{\mu}$&
&
$-2\eta_{3}^{2}\frac{1}{\sqrt{a}}$&
&
$\left(-\frac{1}{2}\right)\frac{1}{\cos^{2}\theta_{W}}=-0.64$&
\tabularnewline
&
&
&
&
&
\tabularnewline
$H_{2}W_{\mu}^{+}W^{\mu}$&
&
$2\left(\eta_{2}^{2}-\eta_{3}^{2}\right)\frac{1}{\sqrt{a}}$&
&
$-\frac{1}{\sqrt{a}}\tan^{2}\theta_{W}=-0.28$&
\tabularnewline
&
&
&
&
&
\tabularnewline
$H_{2}Z_{\mu}Z^{\mu}$&
&
$\frac{\left(\eta_{2}^{2}-\eta_{3}^{2}\right)}{\cos^{2}\theta_{W}}\frac{1}{\sqrt{a}}$&
&
$\left(-\frac{1}{2}\right)\frac{\tan^{2}\theta_{W}}{\cos^{2}\theta_{W}}=-0.37$&
\tabularnewline
&
&
&
&
&
\tabularnewline
$H_{2}Z_{\mu}^{\prime}Z^{\prime\mu}$&
&
$\frac{\left[\left(1-2\sin^{2}\theta_{W}\right)^{2}\eta_{2}^{2}-\eta_{3}^{2}\right]}{\left(3-4\sin^{2}\theta_{W}\right)\cos^{2}\theta_{W}}\frac{1}{\sqrt{a}}$&
&
$\left(-\frac{1}{2}\right)\frac{\tan^{2}\theta_{W}\left(3-6\sin^{2}\theta_{W}+4\sin^{4}\theta_{W}\right)}{\left(3-4\sin^{2}\theta_{W}\right)}=-0.12$&
\tabularnewline
&
&
&
&
&
\tabularnewline
$H_{3}Y_{\mu}^{+}Y^{\mu}$&
&
$2\eta_{2}\eta_{3}\frac{1}{\sqrt{a}}$&
&
$\frac{\sqrt{1-2\sin^{2}\theta_{W}}}{2\cos^{2}\theta_{W}}=0.47$&
\tabularnewline
&
&
&
&
&
\tabularnewline
$H_{3}X_{\mu}^{+}X^{\mu}$&
&
$2\eta_{2}\eta_{3}\frac{1}{\sqrt{a}}$&
&
$\frac{\sqrt{1-2\sin^{2}\theta_{W}}}{2\cos^{2}\theta_{W}}=0.47$&
\tabularnewline
&
&
&
&
&
\tabularnewline
$H_{3}W_{\mu}^{+}W^{\mu}$&
&
$4\eta_{2}\eta_{3}\frac{1}{\sqrt{a}}$&
&
$\frac{\sqrt{1-2\sin^{2}\theta_{W}}}{\cos^{2}\theta_{W}}=0.95$&
\tabularnewline
&
&
&
&
&
\tabularnewline
$H_{3}Z_{\mu}Z^{\mu}$&
&
$\frac{2\eta_{2}\eta_{3}}{\cos^{2}\theta_{W}}\frac{1}{\sqrt{a}}$&
&
$\frac{\sqrt{1-2\sin^{2}\theta_{W}}}{2\cos^{4}\theta_{W}}=0.61$&
\tabularnewline
&
&
&
&
&
\tabularnewline
$H_{3}Z_{\mu}^{\prime}Z^{\prime\mu}$&
&
$\frac{\eta_{2}\eta_{3}\left[\left(1-2\sin^{2}\theta_{W}\right)^{2}+1\right]}{\left(3-4\sin^{2}\theta_{W}\right)\cos^{2}\theta_{W}}\frac{1}{\sqrt{a}}$&
&
$\frac{\left(1-2\sin^{2}\theta_{W}+2\sin^{4}\theta_{W}\right)\sqrt{1-2\sin^{2}\theta_{W}}}{2\left(3-4\sin^{2}\theta_{W}\right)\cos^{4}\theta_{W}}=0.19$&
\tabularnewline
\hline
\end{tabular}
\end{table}

\subsubsection{Higgs decay rates}

The most general decay scenario is the one in which each Higgs comes
out heavier than double mass of the heaviest boson to which it couples.
so all channels are kinematically allowed ). 

\begin{equation}
\begin{array}{ccccc}
H_{1}\rightarrow X^{+}X &  & H_{1}\rightarrow Y^{+}Y &  & H_{1}\rightarrow Z^{\prime}Z^{\prime}\\
\\ & H_{2}\rightarrow W^{+}W &  & H_{2}\rightarrow ZZ\\
\\ & H_{3}\rightarrow W^{+}W &  & H_{3}\rightarrow ZZ\end{array}\label{Eq. 37}\end{equation}

The general formula for the partial width of the Higgs decay into
two any gauge bosons is given in the Born approximation (at tree level)
by the well-known formula:

\begin{equation}
\Gamma(H\rightarrow BB)=g_{HBB}^{2}\frac{m(H)^{2}\alpha}{32\pi\sqrt{2}\left\langle \phi\right\rangle ^{2}}\sqrt{1-\frac{4m(B)^{2}}{m(H)^{2}}}\left(4-\frac{16m(B)^{2}}{m(H)^{2}}+\frac{48m(B)^{2}}{m(H)^{4}}\right)\label{Eq. 38}\end{equation}
with $\alpha=1$for neutral bosons and $\alpha=2$ for charged ones
and $B$ denoting any gauge boson in the model. Noting the ratio $x=4M_{W}^{2}/m_{H}^{2},$
the concrete functions can be computed as depending only on the couplings
$g_{HBB}$, ratio $x$ and parameter $a$.

\section{Results and numerical estimates}

\subsection{Plausible scenarios}

Up to this point, our approach has been a pure theoretical exercise
stemming from the fertile soil of the SM. At this moment one can test
some plausible scenarios beyond SM by choosing certain orders of magnitude
for the overall VEV $\left\langle \varphi\right\rangle $. Hence,
some rough estimates are obtained for the resulting phenomenology.
We work out here the case of interest in which $\left\langle \varphi\right\rangle \in(1-10)$
TeV with the three VEVs aligned as: 

\begin{itemize}
\item $<\rho>\in(\sqrt{1-a}\div10\sqrt{1-a})$TeV, 
\item $<\chi>\simeq\left(\sqrt{\frac{\left(1-\tan^{2}\theta_{W}\right)}{2}}\right)\left\langle \varphi\right\rangle _{SM}=147.6$GeV 
\item $<\phi>\simeq\sqrt{\frac{1}{2\cos^{2}\theta_{W}}}\left\langle \varphi\right\rangle _{SM}=197$GeV
\end{itemize}
implying $a\in$$\left(0.0006-0.06\right)$ as it results from $\sqrt{a}\left\langle \varphi\right\rangle =\left\langle \varphi\right\rangle _{SM}$
in order to ensure $m(W)=80.4$GeV and $m(Z)=91.1$GeV. 

Before entering the discussion of the Higgs phenomenology and its
restrictions, let's estimate the implications of some verified phenomenological
aspects \cite{key-45}. For instance, the ''wrong muon decay'' gives
at a 98\% CL the result

\begin{equation}
R=\frac{\Gamma(\mu^{-}\rightarrow e^{-}\bar{\nu_{\mu}}\nu_{e})}{\Gamma(\mu^{-}\rightarrow e^{-}\bar{\nu_{e}}\nu_{\mu})}=\left(\frac{M_{W}}{M_{Y}}\right)^{4}\leq1.2\%\label{Eq. 39}\end{equation}
Hence $M_{Y}\geq240$GeV or equivalently - in our approach - to $a\leq0.123$,
which is already fulfilled. 

With the allowed range of the parameter $a$, one can compute the
allowed domain for boson masses. These are, at the presumed breaking
scales, those presented in Table 2. 

\begin{table}

\caption{Masses of the gauge bosons in 331RHN model}

\begin{tabular}{ll|cc|cc|cc}
\hline 
Mass&
&
at $\left\langle \varphi\right\rangle =1$TeV&
&
at $\left\langle \varphi\right\rangle =5$TeV&
&
at$\left\langle \varphi\right\rangle =10$TeV&
\tabularnewline
&
&
$a=0.06$&
&
$a=0.0024$&
&
$a=0.0006$&
\tabularnewline
\hline
\hline 
&
&
&
&
&
&
&
\tabularnewline
$m(Y)$&
&
$321.8$GeV&
&
$1.64$TeV&
&
$3.28$TeV&
\tabularnewline
&
&
&
&
&
&
&
\tabularnewline
$m(X)$&
&
$324.7$GeV&
&
$1.64$TeV&
&
$3.28$TeV&
\tabularnewline
&
&
&
&
&
&
&
\tabularnewline
$m(Z^{\prime})$&
&
$389.2$GeV&
&
$1.99$TeV&
&
$3.98$TeV&
\tabularnewline
&
&
&
&
&
&
&
\tabularnewline
\hline
\end{tabular}
\end{table}

\subsection{Perturbativity}

Now, in order to keep the Higgs phenomenology in the perturbative
regime, the numerical values of the couplings in Table 1 must not
overcome those in SM. That obviously happens, since each of them (except
for those involving $H_{1}$) exhibit couplings less than those in
SM, as one can read from the last column of Table 1. For $H_{1}$
that requirement enforces a lower bound on parameter $a$. For the
considered domain of the breaking scale, the lower bound is $a\geq0.0027$
in the case $\left\langle \varphi\right\rangle =1$TeV, $a\geq0.00052$
in the case $\left\langle \varphi\right\rangle =5$TeV, and respectively
$a\geq0.00013$ in the case $\left\langle \varphi\right\rangle =10$TeV,
that are automatically satisfied. So, there are no problems with perturbativity
due to HBB couplings or HHBB.

By inspecting trilinear and quartic couplings of the Higgs bosons
- $\mathbf{g\mathbf{(HHH)}}$ and $g\mathbf{(HHHH)}$ from Eqs. (\ref{Eq. 18})
and (\ref{Eq. 19}) - one can derive an upper bound on their masses,
if they are set up to keep perturbativity. That is, the couplings
must also remain below 1 at the considered breaking scale.

\begin{equation}
g\mathbf{(HHH)=4\lambda\eta_{i}\left\langle \varphi\right\rangle }\,,\quad g(HHHH)=\lambda\,,\label{Eq. 40}\end{equation}

Consequently, one obtains $\lambda<1/4$. Assuming that $H_{3}$ is
the SM Higgs boson, its experimental constraints \cite{key-46,key-47}
impose $m_{3}\geq114.4$GeV\cite{key-45}. If we take the upper limit
for $\lambda=1/4$ , then in order to get a safe behavior concerning
perturbativity, the Higgs masses become: 

\begin{equation}
m_{1}\cong\frac{1}{\sqrt{a}}\sqrt{\left(1-a\right)}\left\langle \varphi\right\rangle _{SM}\label{Eq. 41}\end{equation}

\begin{equation}
m_{2}\cong\frac{1}{2}\sqrt{\frac{1}{2}\left(1-\tan^{2}\theta_{W}\right)}\left\langle \varphi\right\rangle _{SM}\label{Eq. 42}\end{equation}

\begin{equation}
m_{3}\cong\frac{1}{2}\sqrt{\frac{1}{2}\left(\frac{4\cos^{2}\theta_{W}-1}{\cos^{2}\theta_{W}}\right)}\left\langle \varphi\right\rangle _{SM}\label{Eq. 43}\end{equation}

Numerical estimates yield precisely $m_{2}=73.44$GeV and $m_{3}=143.25$GeV.
The new Higgs develops distinct masses, in the following cases: $m_{1}=973.7$GeV
when $\left\langle \varphi\right\rangle =1$TeV, $m_{1}=5.01$TeV
when $\left\langle \varphi\right\rangle =5$TeV and $m_{1}=10.03$TeV
when $\left\langle \varphi\right\rangle =10$TeV respectively.

This state of affairs leads - as expected - to the conclusion that
$H_{2},H_{3}\rightarrow Z^{\prime}Z^{\prime}$, $H_{2},H_{3}\rightarrow YY$,
and $H_{2},H_{3}\rightarrow XX{}$ are completely forbidden. In addition,
neither $H_{2},H_{3}\rightarrow ZZ$ nor $H_{2},H_{3}\rightarrow W^{+}W^{-}$
occur. Therefore, no decay event with regard to those two ''lighter''
Higgs to vector bosons is expected to be observed.

\subsection{Loop corrections}

Furthermore, it is interesting to investigate if such Higgs bosons
do alter somehow - by means of radiative corrections - the parameter
$\rho$, the masses of the SM bosons $W$ and $Z$. We restrict ourselves
here to inspect the 1-loop corrections. First of all, one notices
that the biggest Higgs $H_{1}$does not interact with SM bosons, so
its contribution to 1-loop corrections will be identical zero. The
other two Higgs have slightly different couplings to SM boons, so
their contributions will be different, since $m_{3}>M_{W}$and $m_{2}<M_{W}$.
The formula giving the 1-loop contribution to $\rho$ of a neutral
scalar field interacting with $W$ and $Z$ was computed decades ago
in \cite{key-48} - \cite{key-53}. It is:

\begin{equation}
\left(\Delta\rho\right)^{1-loop}=-\frac{3G_{F}M_{W}^{2}}{8\sqrt{2}\pi^{2}}\left[\left(-0.28\right)f\left(\frac{m_{2}^{2}}{M_{W}^{2}}\right)+\left(0.95\right)f\left(\frac{m_{3}^{2}}{M_{W}^{2}}\right)\right]\label{Eq. 44}\end{equation}
where we introduced the actual couplings $g\left(H_{2}WW\right)=-0.28\times(2m_{W}^{2}/\left\langle \varphi\right\rangle _{SM}$
and $g\left(H_{3}WW\right)=0.95\times(2m_{W}^{2}/\left\langle \varphi\right\rangle _{SM}$.
The function $f$ is

\begin{equation}
f\left(x\right)=x\left[\frac{\ln c_{W}^{2}-\ln x}{\ln c_{W}^{2}-x}+\frac{\ln x}{\ln c_{W}^{2}\left(1-x\right)}\right]\label{Eq. 45}\end{equation}

Assuming the above order of magnitude for the Higgs masses, the 1-loop
radiative correction to $\rho$ parameter due to Higgs contribution
yields: $0.008$. Furthermore, if one wants to calculate the 1-loop
contribution of the Higgs sector to the mass of the $W$ boson, one
can use the celebrated formula obtained in Refs. \cite{key-54} -
\cite{key-57}

\begin{equation}
M_{W}^{2}\left(1-\frac{M_{W}^{2}}{M_{Z}^{2}}\right)=\frac{\pi\alpha}{\sqrt{2}G_{F}}\left(1+\Delta r\right)\label{Eq. 46}\end{equation}

with $\left(\Delta r\right)^{1-loop}$ as in Refs. \cite{key-54}
- \cite{key-60} but taking into consideration our specific couplings:

\begin{equation}
\left(\Delta r\right)^{1-loop}\simeq\frac{G_{F}M_{W}^{2}}{8\sqrt{2}\pi^{2}}\frac{11}{3}\left[\left(-0.28\right)\left(\log\frac{m_{2}^{2}}{M_{W}^{2}}-\frac{5}{6}\right)+\left(0.95\right)\left(\log\frac{m_{3}^{2}}{M_{W}^{2}}-\frac{5}{6}\right)\right]\label{Eq. 47}\end{equation}

This yields, in the case of interest here, a negligible amount $\left(\Delta r\right)^{1-loop}\simeq0.0009$.

\subsection{Higgs production}

On the experimental level, at the LHC the Higgs ''hunting'' is currently
in the run and has raised big expectations. In the 331RHN model there
are three distinct kinds of producing the SM-like Higgs boson. The
processes to be watched are in order: (a) $pp\rightarrow ZH_{3}$,
(b) , $pp\rightarrow Z'H_{3}$ and respectively (c) $pp\rightarrow Z'$
and then following the decay modes of $Z'$ such as $Z'\rightarrow H_{3}B$
(where $B$ denotes a neutral gauge bosons). Some numerical analyses
have been performed for such processes in Ref. \cite{key-43} in slightly
different scenarios, therein assuming the exotic quarks with masses
similar to the heaviest Higgs ($M(Q)\simeq m_{1}$). However, roughly
speaking, the (c) way gives less hope in our scenario since the resulting
total width of the $Z'$ seems to be greater than that in Ref. \cite{key-43},
as our $M_{Z'}$is significantly greater when $\left\langle \varphi\right\rangle $
goes to $10$TeV, so consequently the branching ratio $\Gamma(Z'\rightarrow HZ)/\Gamma(Z'\rightarrow all)$
diminishes. At the same time, the (b) route can be ignored, as the
total cross section of such $pp$ processes is negligible too, even
for lighter $Z'$(Fig.6 in Ref. \cite{key-43} proves this in the
case $M_{Z'}\in1-2$TeV), while our $M_{Z'}$ reaches even $3.9$TeV).
So, the remaining process to be thoroughly investigated with numerical
accuracy is the Higgs production via $Z$ boson exchange in $pp$
collisions and it will be performed in a future work. However, from
Fig.4 in Ref. \cite{key-43} one can read a rough estimate for our
SM-like Higgs bosons. This indicts a total cross section of about
$1$pb from $Z$ exchange, and at most $10^{-3}$pb from $Z'$ exchange,
if we assume an average $2$TeV mass for the heavy $Z'$. Yet, if
$M_{Z'}$is greater, the (c) channel's cross section diminishes even
more. Therefore, (a) remains the most relevant process to be sought-after
at the LHC and to be work out in a separate paper.

\section{Concluding remarks}

We have discussed here the Higgs sector of a 331RHN gauge model and
suggested a plausible scenario supplied by an overall breaking scale$\left\langle \varphi\right\rangle \in1-10$
TeV. Our work primarily proves that the particular method conceived
by Cot\u{a}escu and developed by the author in previous papers can
be successfully accommodated with the traditional approach in the
literature, by simply redefining the scalar multiplets, so that instead
of one surviving Higgs field there are three such physical fields
in the end. Yet, the advantage of tuning a single free parameter is
kept here and it is exploited in order to make some phenomenological
predictions such as: boson masses $M_{X}=M_{X}(a)$, $M_{Y}=M_{Y}(a)$
and $M_{Z'}=M_{Z'}(a)$ and Higgs masses $m_{1}\cong\left\langle \varphi\right\rangle $TeV,
$m_{2}=73$GeV, $m_{3}=143$GeV - all independently of the free parameter
$a$, while the SM phenomenology is entirely recovered. It remains
to be analyzed the Higgs contributions in higher loops diagrams (of
$\rho$parameter and SM bosons mass), in order to fulfill the renormalizability
requirement for such theories and work out the details of the Higgs
production from $Z$ exchange processes.


\begin{thebibliography}{10}
\bibitem{key-1}S. Weinberg, \emph{Phys. Rev. Lett.} \textbf{19}, 1264 (1967). 
\bibitem{key-2}S. L. Glashow. \emph{Nucl. Phys.} \textbf{20}, 579 (1961).
\bibitem{key-3}A. Salam, in \emph{''Elementary Particle Theory''}: Relativistic
Groups in Analyticity (Nobel Symposium No.8), edited by N. Svartholm,
Almqvist and Wiknell - Stockholm 1968, p.367.
\bibitem{key-4}A. Strumia and F. Vissani, arXiv: hep-ph/0606054 v3. 
\bibitem{key-5}P. W. Higgs, \emph{Phys. Rev. Lett.} \textbf{13}, 508 (1964). 
\bibitem{key-6}F. Englert and R. Brout, \emph{Phys. Rev. Lett.} \textbf{13}, 321
(1964).
\bibitem{key-7}G. S. Guralnik, C. R. Hagen and T. Kibble, \emph{Phys. Rev. Lett.}
\textbf{13}, 585 (1965).
\bibitem{key-8}P. W. Higgs, \emph{Phys. Rev.} \textbf{145}, 145 (1966).
\bibitem{key-9}T. Kibble, \emph{Phys. Rev.} \textbf{155}, 1554 (1967).
\bibitem{key-10}P. H. Frampton, \emph{Phys. Rev. Lett.} \textbf{69}, 2889 (1992). 
\bibitem{key-11}F. Pisano and V. Pleitez, \emph{Phys. Rev.} \emph{D} \textbf{46},
410 (1992).
\bibitem{key-12}R. Foot, H. N. Long and T. A. Tran, \emph{Phys. Rev.} \emph{D} \textbf{50},
R34 (1994).
\bibitem{key-13}H. N. Long, \emph{Phys. Rev.} \emph{D} \textbf{53}, 437 (1996).
\bibitem{key-14}H. N. Long, \emph{Phys. Rev.} \emph{D} \textbf{54}, 4691 (1996).
\bibitem{key-15}W. A. Ponce, J. B. Florez and L. A. Sanchez, \emph{Int. J. Mod. Phys.
A} \textbf{17}, 643 (2002).
\bibitem{key-16}L. A. Sanchez, W. A. Ponce and R. Martinez, \emph{Phys. Rev.} \emph{D}
\textbf{64}, 075013 (2001).
\bibitem{key-17}R. A. Diaz, R. Martinez and F. Ochoa, \emph{Phys. Rev.} \emph{D} \textbf{72},
035018 (2005).
\bibitem{key-18}I. I. Cotaescu, \emph{Int. J. Mod. Phys. Rev.} \emph{A} \textbf{12},
1483(1997).
\bibitem{key-19}A. Palcu, \emph{Mod. Phys. Lett.} \emph{A} \textbf{21}, 1203 (2006).
\bibitem{key-20}A. Palcu, \emph{Mod. Phys. Lett.} \emph{A} \textbf{23}, 387 (2008).
\bibitem{key-21}H. N. Long and T. Inami, \emph{Phys. Rev. D} \textbf{61}, 075002 (2000).
\bibitem{key-22}G. Tavares-Velasco and J. J. Toscano, \emph{Phys. Rev.} \emph{D} \textbf{70},
053006 (2004).
\bibitem{key-23}A. G. Dias, J. C . Montero and V. Pleitez, \emph{Phys. Rev.} \emph{D}
\textbf{73}, 113004 (2006).
\bibitem{key-24}A. Doff, C. A. de S. Pires and P. S. Rodrigues da Silva, \emph{Phys.
Rev.} \emph{D} \textbf{74}, 015014 (2006).
\bibitem{key-25}A. Carcamo, R. Martinez and F. Ochoa, \emph{Phys. Rev.} \emph{D} \textbf{73},
035007 (2006).
\bibitem{key-26}F. Ramirez-Zavaleta, G. Tavares-Velasco and J. J. Toscano, \emph{Phys.
Rev.} \emph{D} \textbf{75}, 075008 (2007).
\bibitem{key-27}E. Ramirez-Barreto, Y. A. Coutinho and J. Sa Borges, \emph{Eur. Phys.
J.} \emph{C} \textbf{50}, 909 (2007).
\bibitem{key-28}D. Cogollo, H. Diniz, C. A. de S. Pires and P. S. Rodrigues da Silva,
\emph{Mod. Phys. Lett.} \emph{A} \textbf{23}, 3405 (2008).
\bibitem{key-29}J. M. Cabarcas, D. Gomez Dumm and R. Martinez, \emph{Eur. Phys. J.}
\emph{C} \textbf{58}, 569 (2008).
\bibitem{key-30}R. H. Benavides, T. Giraldo and W. A. Ponce, \emph{Phys. Rev.} \emph{D}
\textbf{80}, 113009 (2009).
\bibitem{key-31}Y.-B. Liu and X.-L. Wang, \emph{Mod. Phys. Lett. A} \textbf{24}, 1307
(2009).
\bibitem{key-32}P. V. Dong, L. T. Hue, H. N. Long and D. V. Soa, \emph{Phys. Rev.}
\emph{D} \textbf{81}, 053004 (2010).
\bibitem{key-33}A. G. Dias, C. A. de S. Pires and P. S. Roridgues da Silva, \emph{Phys.
Lett. B} \textbf{628}, 85(2005).
\bibitem{key-34}D. Chang and H. N. Long, \emph{Phys. Rev.} \emph{D} \textbf{73}, 053006
(2006).
\bibitem{key-35}A. Palcu, \emph{Mod. Phys. Lett.} \emph{A} \textbf{21}, 2027 (2006).
\bibitem{key-36}A. Palcu, \emph{Mod. Phys. Lett.} \emph{A} \textbf{21}, 2591 (2006).
\bibitem{key-37}A. Palcu, \emph{Mod. Phys. Lett.} \emph{A} \textbf{22}, 939 (2007).
\bibitem{key-38}D. Cogollo, H. Diniz, C. A. de S. Pires and P. S. Rodrigues da Silva,
\emph{Eur. Phys. J. C} \textbf{58}, 455 (2008).
\bibitem{key-39}P. V. Dong and H. N. Long, \emph{Phys. Rev.} \emph{D} \textbf{77},
057302 (2008).
\bibitem{key-40}D. Cogollo, H. Diniz and C. A. de S. Pires, \emph{Phys. Lett B} \textbf{677},
338 (2009).
\bibitem{key-41}H. N. Long, \emph{Mod. Phys. Lett.} \emph{A} \textbf{13}, 1865 (1998).
\bibitem{key-42}R. A. Diaz, R. Martinez and F. Ochoa, \emph{Phys. Rev.} \emph{D} \textbf{69},
095009 (2004). 
\bibitem{key-43}L. D. Ninh and H. N. Long, \emph{Phys. Rev.} \emph{D} \textbf{72},
075004 (2005). 
\bibitem{key-44}Y. Giraldo and W. A. Ponce, arXiv: 1107.3260 {[}hep-ph{]}.
\bibitem{key-45}K. Nakamura et al. (PDG), \emph{J. Phys. G} \textbf{37}, 075021 (2010).
\bibitem{key-46}A. Djouadi, \emph{Phys. Rept.} \textbf{457}, 1-216 (2008).
\bibitem{key-47}A. Djouadi, \emph{Phys. Rept.} \textbf{459}, 1-241 (2008).
\bibitem{key-48}M. Veltman, \emph{Acta Phys. Pol.} \emph{B} \textbf{8}, 475 (1977).
\bibitem{key-49}T. Appelquist and C. Bernard, \emph{Phys. Rev. D} \textbf{22}, , 200
(1980).
\bibitem{key-50}A. Longhitano, \emph{Nucl. Phys. B} \textbf{188}, 118 (1981).
\bibitem{key-51}M. Einhorn and J. Wudka, \emph{Phys. Rev. D} \textbf{39}, 2758 (1989).
\bibitem{key-52}J. J. van der Bij and M. Veltman, \emph{Nucl. Phys. B} \textbf{231},
205 (1985).
\bibitem{key-53}R. Boghezal, J. B. Tausk and J. J. van der Bij, \emph{Nucl. Phys.
B} \textbf{713}, 278 (2005).
\bibitem{key-54}A. Sirlin, \emph{Phys. Rev. D} \textbf{22}, 971 (1980).
\bibitem{key-55}W. Marciano and A. Sirlin, \emph{Phys. Rev. D} \textbf{22}, 2605 (1980).
\bibitem{key-56}W. Marciano and A. Sirlin, \emph{Phys. Rev. D} \textbf{29}, 945 (1984).
\bibitem{key-57}A. Sirlin and W. Marciano, \emph{Nucl. Phys. B} \textbf{189}, 442
(1981).
\bibitem{key-58}W. Marciano, \emph{Phys. Rev. D} \textbf{20}, 274 (1979).
\bibitem{key-59}W. Marciano and A. Sirlin, \emph{Phys. Rev. Lett.} 46, 163 (1981).
\bibitem{key-60}W. Marciano, S. Srantakos and A. Sirlin, \emph{Nucl. Phys. B} \textbf{217},
64 (1983).\end{thebibliography}
\end{document}